\documentclass[prb,aps,twocolumn,floatfix]{revtex4}
\pdfoutput 1
\usepackage{graphicx,epsfig}
\usepackage{times}
\usepackage{amsmath,amssymb}

\newlength\figurewidth
\begin{document}

\title{Surface plasmon polariton modes in a single-crystal Au nanoresonator fabricated using focused-ion-beam milling}

\author{E. J. R. Vesseur}
\email{vesseur@amolf.nl}
\author{R. de Waele}
\author{A. Polman}
\affiliation{Center for Nanophotonics, FOM Institute for Atomic and Molecular Physics (AMOLF), Kruislaan~407, 1098~SJ~Amsterdam, The~Netherlands}

\author{H. J. Lezec}
 \altaffiliation[present address: ]{Center for Nanoscale Science and Technology, NIST, Gaithersburg, MD, USA}
\author{H. A. Atwater}
\affiliation{Thomas J. Watson Laboratories of Applied Physics, California Institute of Technology, MS~128-95, Pasadena, California~91125}

\author{F. J. Garc\'ia de Abajo}
\affiliation{Instituto de \'Optica - CSIC, Serrano~121, 28006, Madrid, Spain}

\begin{abstract}
We use focused-ion-beam milling of a single-crystal Au surface to fabricate a 590-nm-long linear ridge that acts as a surface plasmon nanoresonator. Cathodoluminescence imaging spectroscopy is then used to excite and image surface plasmons on the ridge.
Principal component analysis reveals distinct plasmonic modes, which proves confinement of surface-plasmon oscillations to the ridge.
Boundary-element-method calculations confirm that a linear ridge is able to support highly-localized surface-plasmon modes (mode diameter $< 100$~nm).
The results demonstrate that focused-ion-beam milling can be used in rapid prototyping of nanoscale single-crystal plasmonic components.
\end{abstract}

\maketitle

Surface plasmon polaritons (SPPs) are electromagnetic waves confined to a metal-dielectric interface. As shown in recent experiments, SPPs can be manipulated using waveguides \cite{rs04_krenn,nat03_barnes} and resonators \cite{prl06_miyazaki,prl05_ditlbacher}. Surface plasmon polaritons hold promise for application in sensing, photovoltaics, telecommunications, and opto-electronic circuit integration, due to their ability to concentrate and guide electromagnetic energy at the nanoscale.

Fabrication of components that guide and confine SPPs involves structuring of metals, typically using methods such as electron beam lithography, nano-imprint lithography, self-assembly or templating techniques. These methods offer high spatial resolution, but require complex multi-step processing. Moreover, the metal films, most often obtained by thermal evaporation, typically have a polycrystalline structure. Grain boundaries and surface roughness in polycrystalline films are known to cause undesired scattering of SPPs. 

A single-step method to structure metals for plasmonic applications, that is gaining widespread acceptance, is focused-ion-beam (FIB) milling \cite{sc02_lezec,mrs07_volkert}. In a typical FIB system, Ga$^+$ ions are extracted from a liquid-metal ion source, accelerated to 30~keV, and focused by an electrostatic lens system to a spot size with diameter as small as 5 to 10~nm.

In this Letter we show how FIB milling of a single-crystal Au substrate can be used for highly-reproducible, maskless fabrication of a smooth plasmonic resonator, with minimum lateral dimensions of 50~nm, and surface roughness on the scale of only a few nm. 
We use cathodoluminescence (CL) imaging spectroscopy to generate SPPs\cite{pr57_ritchie,javier_unknown} and image resonant modes within the metal nanostructures. The data demonstrate that FIB is an ideal tool for fabrication of nanoscale plasmonic components, in particular when single-crystal metal substrates are employed. Our conclusions are supported by boundary-element-method (BEM) calculations of the local fields at the metal nanostructures\cite{prl98_garcia-de-abajo_howie,prb99_garcia-de-abajo}. 

\begin{figure}
\includegraphics[width=\figurewidth]{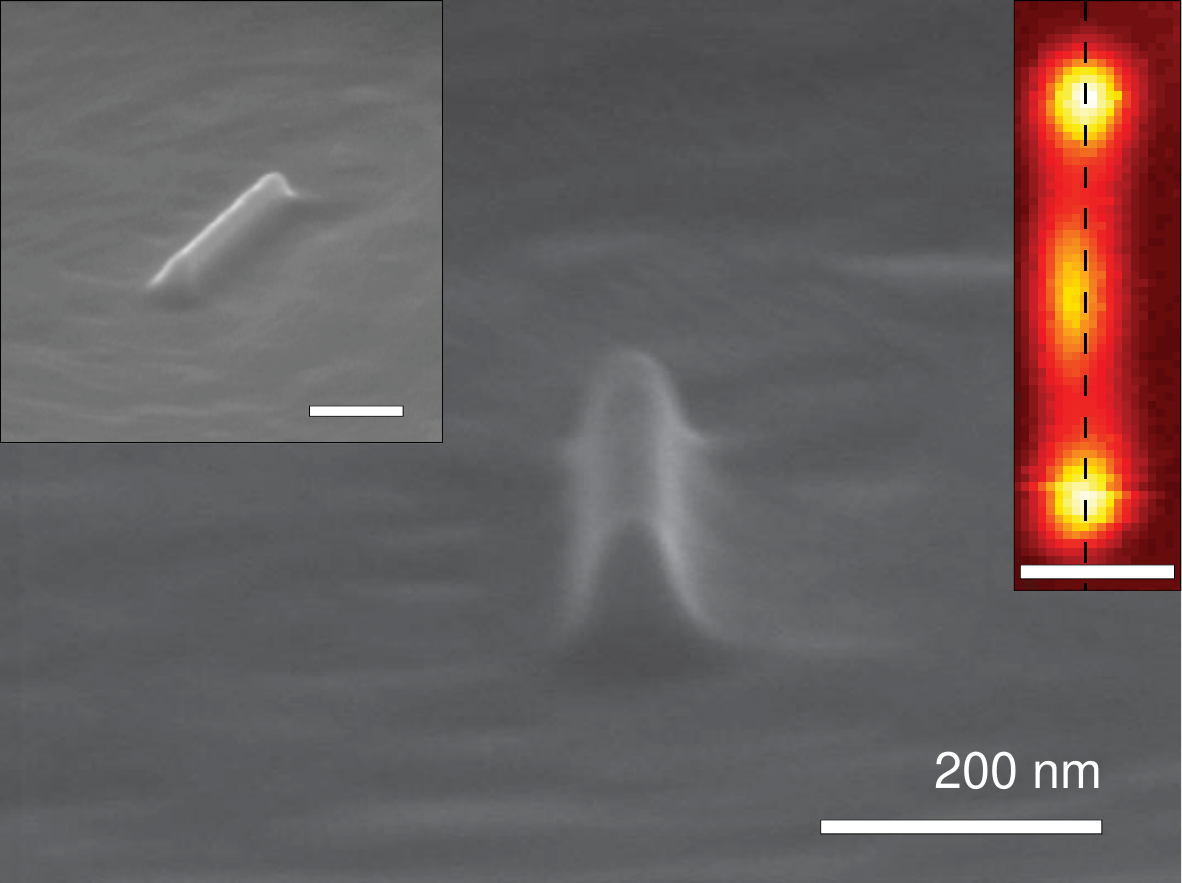}
\caption{Scanning electron micrograph of a 590-nm-long linear ridge fabricated in a single-crystal Au substrate using focused-ion-beam milling. The ridge is approximately 95~nm high and 80~nm wide.
The left inset shows the ridge from a different angle. 
The inset on the right shows the light emission from the ridge at 585~nm wavelength upon irradiation with a 30~kV electron beam, as function of electron beam position. All scale bars are 200~nm.}\label{sem}
\end{figure}

Nano-plasmonic device fabrication consisted of focused-ion-beam milling a polished $<$111$>$ single-crystal Au substrate, grown using the Czochralski process.
A 10~pA, 30~keV Ga beam from an FEI Nova 600 dual-beam workstation was rastered in 40 passes over a 4 $\times$ 5 $\mu$m area, using a 1000 pixel $\times$ 1250 pixel grid using a variable dwell time per pixel. Subtractive milling left behind a linear ridge, approximately 590-nm long, 95-nm high and 80-nm wide. Fabrication of such a ridge takes roughly one minute and is very reproducible. Resulting ridge roughness is on the order of a few nm, which is approcimately 10 times smoother than what can be achieved using polycrystalline metal substrates.

Figure~\ref{sem} shows scanning-electron-microscope (SEM) images of the ridge, taken at large azimuthal angle with respect to the normal. Views of the structure at two different polar angles are presented, respectively along (main panel), and several degrees off (inset) the ridge axis. 
The images reveal rounded features at the top of the ridge and at the base. We attribute this rounding to redeposition of Au during milling as well as the finite diameter of the ion beam ($\sim$10~nm). The features visible in the milled area around the ridge are attributed to roughness that was initially present on the substrate, and enhanced under ion milling.

CL measurements were done using a FEI XL30 SEM equipped with a parabolic mirror that collects (with a solid angle of 1.4$\pi$~sr) light generated when the electron beam impinges on the sample. The mirror directs the CL emission into a spectrometer that is equipped with a CCD array detector. Spectral data are corrected for system response. The electron beam is scanned across the sample and for every position of the electron beam a spectrum is collected. 

\begin{figure}
\includegraphics[width=\figurewidth]{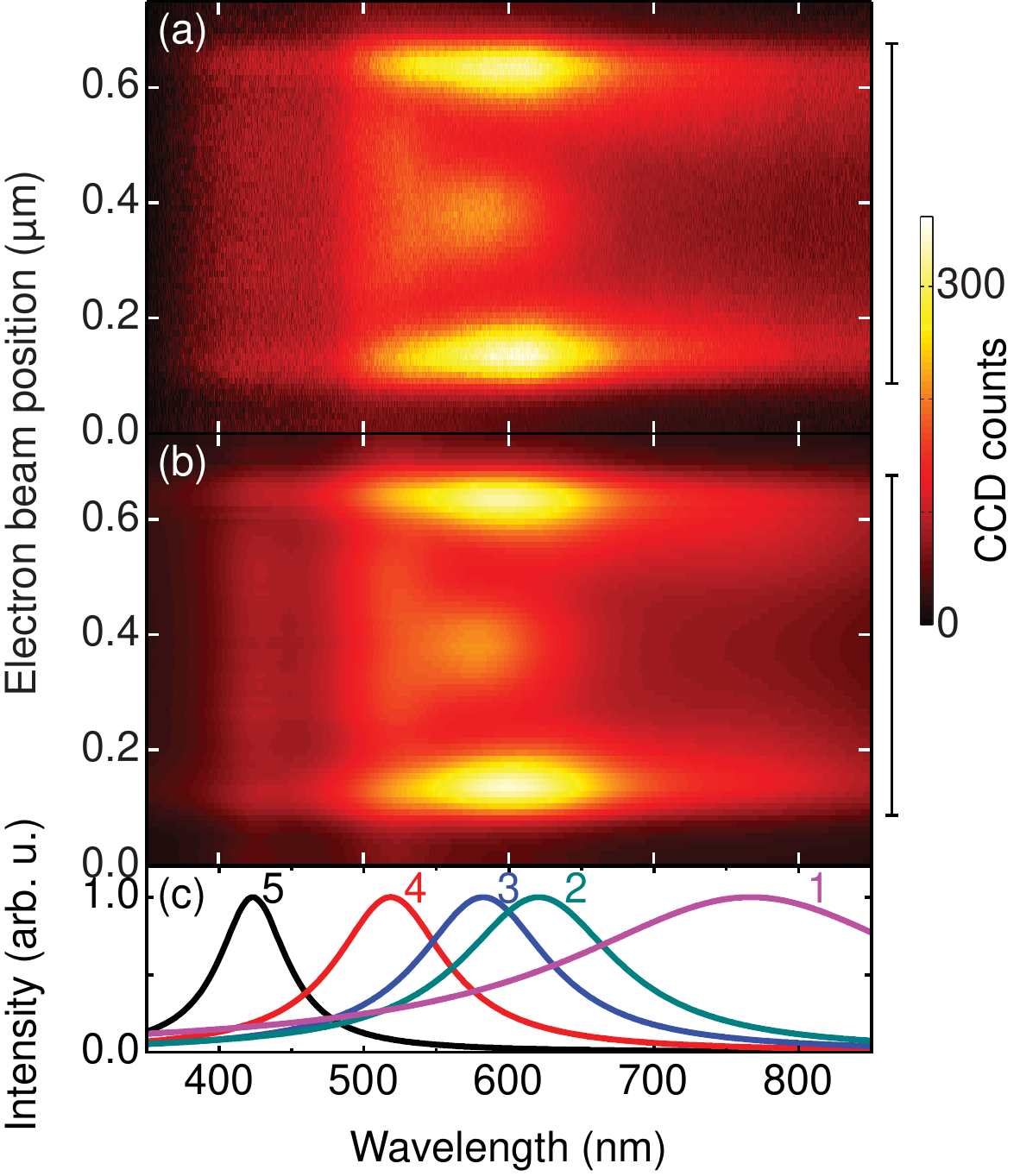}
\caption{(a) Cathodoluminescence emission from a single-crystal Au ridge as function of electron beam position and wavelength, measured along the major axis of the ridge (see right inset of Fig.~\ref{sem}). The ridge position is indicated by the bar to the right of the plot. A broad spectrum is emitted when the electron beam dwells on the ridge ends, while the signal from the ridge center is more sharply wavelength dependent.
(b) Reconstruction of the data from a fit using a modeling factor-analysis method with a Lorentzian shape imposed on five independent spectra, labeled 1 to 5 and plotted in (c). The calculated data set in (b) closely matches the experimental data in (a)}\label{lscan}
\end{figure}

The right inset of Fig.~\ref{sem} shows a CL image of the ridge, acquired at a center wavelength of 585~nm and a spectral bandwidth of 10~nm.
The emission from the ridge is strongly dependent on position: clear maxima are observed at the ends and center of the ridge.
Figure~\ref{lscan}(a) shows spectral line traces (wavelength range: 350--850~nm), along the major axis of the ridge (see dashed line in the inset of Fig.~\ref{sem}) obtained by integration over 3 pixels (total distance: 21~nm) in the lateral direction of the ridge; the ridge position is indicated at the right-hand side of the figure. The data clearly show that a broad spectrum is emitted when the electron beam dwells on the ends of the ridge, while the emission from the center of the ridge is more sharply wavelength dependent. As we will show next, these features are related to resonant geometrical modes of the linear ridge nanoresonator.

As previously shown,\cite{prb01_yamamoto,sia06_yamamoto,nl07_vesseur,nl07_hofmann}, spatial CL images are a direct probe of resonant modes of plasmonic nanostructures.
The measured CL spectrum at any position is assumed to be a linear superposition of multiple principal-mode spectra. We extract the principal modes from the data in Fig.~\ref{lscan}(a) using a factor analysis method\cite{um06_bosman,book91_malinowski}: a two-dimensional data matrix $D_{(x,\lambda)}$ is defined, with one spectral and one spatial dimension, corresponding to the dimensions of Fig.~\ref{lscan}, respectively, as a product of two matrices $D_{(x,\lambda)} = R_{(x,n)} C_{(n,\lambda)}$, one having $n$ spatial profiles as columns and the other having the corresponding $n$ spectra as rows. We find that five principal components form a good representation of the data, so that matrices $R$ and $C$ with $n=5$ can be used to construct the full data set.

\begin{figure}
\includegraphics[width=\figurewidth]{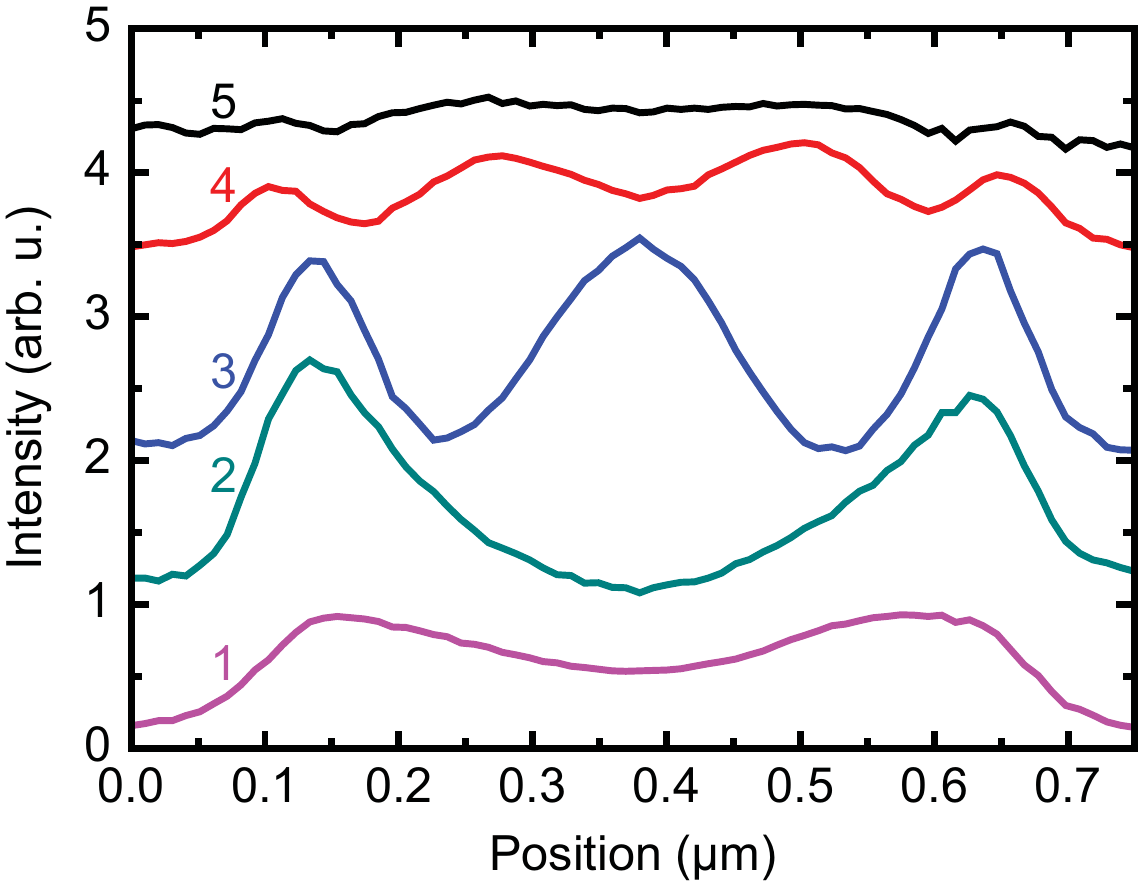}
\caption{Line profiles obtained by a fit of the data in Fig.~\ref{lscan}(a) using the  modeling factor-analysis method of Fig.~\ref{lscan}. Each consecutive profile is shifted by 1 intensity unit for clarity. Numbers labeling each profile correspond to respective spectra of Fig.~\ref{lscan}(c). Profile 3 and 4 show resonant modes of the ridge with one and two antinodes on the ridge, respectively.}\label{profiles}
\end{figure}

We then impose on the mode spectra a Lorentzian shape, parametrized by a peak wavelength and a characteristic width for every spectrum; an initial matrix $C$ is composed of five spectra. The profile matrix $R$ can then be computed by the pseudoinverse~\cite{book91_malinowski} of $C$: $R = D C^{+}$, for which a data set $\hat D = R C$ is then calculated. The parameters for the spectra matrix $C$ are then optimized using a least-squares method by minimizing the difference between the original data set $D$ and the generated $\hat D$. Figure~\ref{lscan}(b) shows the obtained $\hat D$. The result is in excellent agreement with the data, reproducing the spectral shape and intensity at all positions on the ridge. The resulting five spectra from $C$ are plotted in Fig.~\ref{lscan}(c). The peak wavelengths in Fig.~\ref{lscan}(c) range from 425--797~nm. The cavity $Q$ factor, obtained from the width of spectrum 3 is $Q = 5.3$.

Fig.~\ref{profiles} displays the spatial profiles of the five spectra along the ridge, as given by $R$.
Spatial profiles for spectra 3 and 4 are characteristic of resonant modes, with one and two antinodes on the ridge, respectively. The spatial profiles for spectra 1 and 2, at longer wavelengths, show antinodes on the ridge ends. The spatial profile for spectrum 5 shows a relatively constant intensity along the ridge; it may be associated to a higher order mode of which the spatial profile can not be resolved, or transition radiation\cite{pmpes96_yamamoto} that does not have spatial dependence. Note that spectra 2 and 3, which are close together spectrally, have very distinct spatial profiles.

\begin{figure}
\includegraphics[width=\figurewidth]{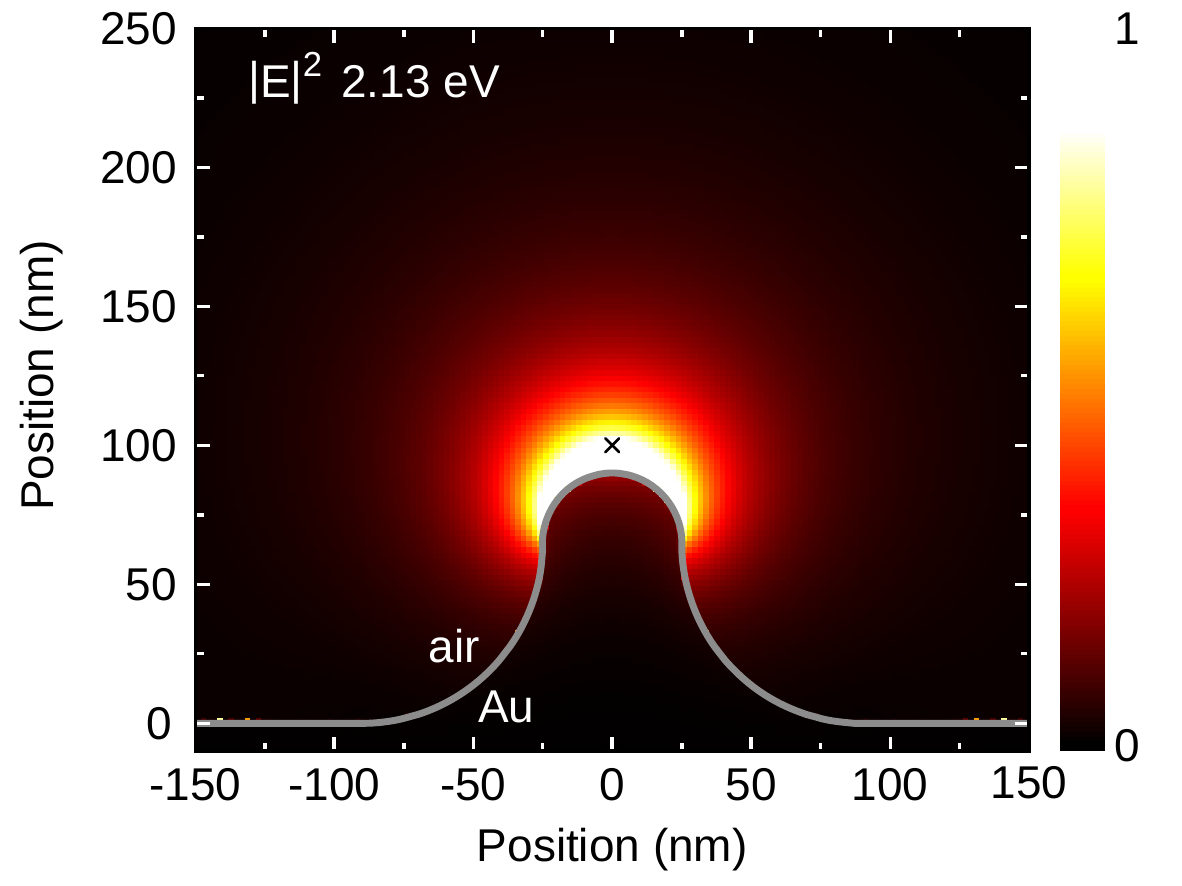}
\caption{Boundary-element-method calculation of the field intensity $|E|^2$ near an infinitely long bell-shaped Au ridge. This calculation, performed at a photon energy 2.13~eV and $k_\parallel = 1.14\ \omega / c$, shows that propagating surface plasmons at this energy are well confined to the top of the ridge and do not couple to the planar-surface plasmons, for which $k_\parallel = 1.07\ \omega / c$.}\label{bem}
\end{figure}

This analysis demonstrates the existence of geometrical plasmon modes along the single-crystal Au ridge that are preferentially excited at antinodes in the modal electric field intensity.
A boundary-element-method (BEM) was used to calculate plasmon modes of a geometry similar to that of the experimental structure. Starting from a shape profile as input, the electric field is expressed in terms of charge and current distribution on the ridge, which are calculated self-consistently to fulfill the field boundary conditions.
We choose a bell-shaped linear ridge of infinite length characterized by two different radii of curvature, as shown by the grey curve in Fig.~\ref{bem}. The shape is similar to that of the ridge imaged in Fig.~\ref{sem}. We first calculated the photonic local density of states (LDOS) in the dielectric just above the top of the ridge (indicated in Fig.~\ref{bem} by the black cross). This was done for a range of energies, separating the LDOS into contributions arising from different spatial frequencies $k_\parallel$ along the ridge.

For example, at an energy of 2.13~eV (corresponding to a free space wavelength of 582~nm, and peak 3 in Fig.~\ref{lscan}(c)), we find an LDOS maximum for $k_\parallel = 1.14\ \omega / c $. This is significantly larger than the corresponding value $k_\parallel = 1.07\ \omega / c$ calculated for SPPs on a planar Au surface (using identical optical constants), consistent with strong mode confinement to the ridge. We investigate the lateral field distribution of the ridge mode by calculating the near field around the ridge excited with a dipole placed above the ridge (also at the position of the cross in Fig.~\ref{bem}). By subtracting the dipole field, we calculate the induced field in the plane normal to the major axis.

Figure~\ref{bem} shows the calculated values of $|E|^2$ around the ridge. Indeed, the figure shows that the intensity is very much confined to the top of the ridge, with a typical lateral confinement in $x$ and $y$ direction well below 100~nm.
From the BEM calculation the plasmon propagation length along an infinitely long ridge is 4~$\mu$m ($Q = 43$). This implies that the measured value $Q = 5.3$, is due to the combined effect of Ohmic losses (possibly enhanced by the presence of implanted Ga from the FIB process) and radiation losses from the ridge ends.

In summary, we have shown that using focused-ion-beam milling, a smooth linear ridge that supports surface plasmon resonances can be fabricated on a single-crystal Au substrate. Spatially-resolved cathodoluminescence spectroscopy shows that surface plasmons generated on this ridge are well confined to the ridge and form distinct geometrical modes. The data are confirmed by boundary-element-method calculations of the plasmon field distribution on the ridge, which show that a propagating surface plasmon is confined to the top of the ridge, with a mode diameter $<$100~nm.
These results show that FIB milling of a single-crystal noble metal surface can be used in rapid prototyping of nanoscale surface plasmon components. 

\begin{acknowledgments}
This work is part of the research program 'Microscopy and modification of nanostructures with focused electron and ion beams'  (MMN) of the 'Stichting voor Fundamenteel Onderzoek der Materie' (FOM), which is financially supported by the 'Nederlandse organisatie voor Wetenschappelijk Onderzoek' (NWO). The MMN program is co-financed by FEI Company. This work is also funded by NANONED, a nanotechnology program of the Dutch Ministry of Economic Affairs. Work at Caltech is financially supported by the Air Force Office of Scientific Research under MURI Grant FA9550-04-1-0434. FJGA wants to thank Prof. Polman and his group for their kind hospitality and support, and acknowledges funding from the EU (STRP-016881-SPANS).
\end{acknowledgments}

\end{document}